\documentclass[aps,prd,preprint,superscriptaddress,showpacs]{revtex4}

\usepackage{graphicx}
\usepackage{amsmath}

\def\be{\begin{equation}}
\def\ee{\end{equation}}
\def\ba{\begin{eqnarray}}
\def\ea{\end{eqnarray}}

\begin{document}

\title{Stability of the normal vacuum in multi-Higgs-doublet models}

\author{A.\ Barroso}
\affiliation{Centro de F\'{\i}sica Te\'{o}rica e Computacional,
    Faculdade de Ci\^{e}ncias,
    Universidade de Lisboa,
    Av.\ Prof.\ Gama Pinto 2,
    1649-003 Lisboa, Portugal}
\author{P.\ M.\ Ferreira}
\affiliation{Centro de F\'{\i}sica Te\'{o}rica e Computacional,
    Faculdade de Ci\^{e}ncias,
    Universidade de Lisboa,
    Av.\ Prof.\ Gama Pinto 2,
    1649-003 Lisboa, Portugal}
\author{R.\ Santos}
\affiliation{Centro de F\'{\i}sica Te\'{o}rica e Computacional,
    Faculdade de Ci\^{e}ncias,
    Universidade de Lisboa,
    Av.\ Prof.\ Gama Pinto 2,
    1649-003 Lisboa, Portugal}
\author{Jo\~{a}o P.\ Silva}
\affiliation{Instituto Superior de Engenharia de Lisboa,
    Rua Conselheiro Em\'{\i}dio Navarro,
    1900 Lisboa, Portugal}
\affiliation{Centro de F\'{\i}sica Te\'{o}rica de Part\'{\i}culas,
    Instituto Superior T\'{e}cnico,
    P-1049-001 Lisboa, Portugal}

\date{\today}

\begin{abstract}
We show that the vacuum structure of a generic multi-Higgs-doublet
model shares several important features with the vacuum structure of
the two and three Higgs-doublet model. In particular, one can still
define the usual charge breaking, spontaneous CP breaking and normal
(charge and CP preserving) stationary points. We analyse
the possibility of charge or spontaneous CP breaking,
by studying the relative depth of the potential in each of the possible
stationary points.
\end{abstract}


\maketitle

\section{\label{sec:intro}Introduction}

Most features of the Standard Model (SM) of electroweak interactions
have been probed to a very high precision. Still, the Higgs sector remains
largely untested and new physics is certainly possible. In particular,
one might have more than one Higgs, as required for example by supersymmetry.
Multi-Higgs models are also appealing for a variety of theoretical reasons
related to CP violation:
i) if the Higgs potential conserves CP, this symmetry could be spontaneously
broken by the vacuum;
ii) if there are three or more Higgs doublets,
then there might be CP violation in the mixing matrix of the charged Higgs;
and iii) the new sources of CP violation have the potential to
explain bariogenesis.

A drawback of these models is that they involve a large number of
parameters. For example, the scalar potential of the most general
two-Higgs-doublet model (2HDM) involves 14 real parameters, while
the potential which explicitly preserves CP involves 10 parameters
(these may be reduced to 11 and 9, respectively, through suitable
basis choices). Given the large number of parameters present in
these models, a variety of methods have been developed in order to
restrict the parameter space, some related to the vacuum structure
of the scalar potential. Recently, some interesting features of the
vacuum structure have been obtained for the particular case of the
2HDM \cite{BFS}. Namely, it was shown that, whenever a normal -
charge and CP conserving - minimum exists in the 2HDM, the global
minimum of that potential is the normal one. Moreover, it was shown
that the depth of the potential at a stationary point that breaks
charge or CP, relative to the normal minimum, is related with the
squared mass matrix of the charged or pseudoscalar Higgs (evaluated
at the normal minimum), respectively. Recent work on these subjects
may be found in~\cite{mani}. In this work we will analyse
how these conclusions may, or may not, be generalised to the case of
a potential with N Higgs doublets.

The paper is organized as follows. In section~\ref{sec:N-Higgs} we
introduce our notation and prove one of our main results: that the
vacuum structure of a generic multi-Higgs-doublet model may be
reduced to vacua involving only two or three doublets. This is
accomplished through a series of basis transformations, for which
the potential is invariant, even if its parameters are not. We
discover that the study of possible charge breaking (CB) vacua is
more easily done in a basis where only three out of the N doublets
have non-vanishing vacuum expectation values (vevs). For CP
violation, the appropriate basis is even simpler: only two doublets
are non-zero. In section~\ref{sec:CB} we compute the values of the
potential at the charge breaking and normal vacua and compare their
values. We study whether it is possible to obtain charge breaking
minima deeper than a normal minimum. In section~\ref{sec:CPV} we
repeat this procedure, but now for CP breaking vacua. We present our
conclusions in section~\ref{sec:conclusions}.
Appendix~\ref{app:CB} provides a basis and gauge independent definition
of the CB vacuum, while appendix~\ref{ap:3hdm}
contains a specific example of a three Higgs-doublet model
for which the CB vacuum lies below the normal vacuum.

\section{\label{sec:N-Higgs}The scalar sector of a generic
N-Higgs-doublet model}

\subsection{The scalar potential}

In this article we follow closely the notation of Refs.~\cite{BS,BLS},
where more details may be found - see also \cite{LS,Gun,Bra,DavHab,GunHab}.
Let us consider a SU(2)$\otimes$U(1) gauge theory with N Higgs-doublets
with the same hypercharge $y=1/2$, denoted by
\be
\Phi_i =
\left(
\begin{array}{c}
\phi^u_i\\
\phi^d_i
\end{array}
\right)
=
\nu_i + \varphi_i
=
\left(
\begin{array}{c}
\nu^u_i\\
\nu^d_i
\end{array}
\right)
+
\left(
\begin{array}{c}
\varphi^u_i\\
\varphi^d_i
\end{array}
\right),
\label{deviate}
\ee
where $\nu_i$ are their vacuum expectation values (vevs),
and $i$ runs from 1 to N. In all that follows, we will use the standard
definition for the electric charge: $Q\,=\,T_3 \, + \, Y$, meaning that all
vevs in the lower components of the doublets are electrically neutral. With
this definition, a vacuum with all upper components of the vevs equal to zero,
$\nu^u_i\,=\,0$, does not break the charge symmetry.

The scalar potential may be written as
\be
V_H = \mu_{ij} (\Phi_i^\dagger \Phi_j) +
\lambda_{ij,kl} (\Phi_i^\dagger \Phi_j) (\Phi_k^\dagger \Phi_l),
\ee
where Hermiticity implies
\ba
\mu_{ij} &=& \mu_{ji}^\ast,
\nonumber\\
\lambda_{ij,kl} \equiv \lambda_{kl,ij} &=& \lambda_{ji,lk}^\ast.
\label{hermiticity_coefficients}
\ea
Under a unitary basis transformation of the Higgs fields,
their kinetic terms remain the same but the coefficients
$\mu_{ij}$ and $\lambda_{ij,kl}$ are transformed in such a
way that the potential remains invariant.
Using Eq.~(\ref{deviate}),
the scalar potential becomes
\ba
V_H
&=&
\mu_{ij}\, (\nu_i^\dagger \nu_j) +
\lambda_{ij,kl}\, (\nu_i^\dagger \nu_j)\, (\nu_k^\dagger \nu_l)
\nonumber\\
&&
+ \nu_i^\dagger\,
\left[ \mu_{ij}  + 2 \lambda_{ij,kl}\, \nu_k^\dagger \nu_l \right]\,
\varphi_j
+
\varphi_i^\dagger\,
\left[ \mu_{ij}  + 2 \lambda_{ij,kl}\, \nu_k^\dagger \nu_l \right]\,
\nu_j
\nonumber\\
&&
+
\mu_{ij}\,  (\varphi_i^\dagger \varphi_j)
+ 2 \lambda_{ij,kl}\,  (\varphi_i^\dagger \varphi_j)\, (\nu_k^\dagger \nu_l)
+ 2 \lambda_{il,kj}\,  (\varphi_i^\dagger \nu_l)\,  (\nu_k^\dagger \varphi_j)
\nonumber\\
&&
\hspace{2.2cm}
+ \lambda_{ij,kl}\,  (\varphi_i^\dagger \nu_j)\,  (\varphi_k^\dagger \nu_l)
+ \lambda_{ij,kl}\,  (\nu_i^\dagger \varphi_j)\,  (\nu_k^\dagger \varphi_l)
\nonumber\\
&&
+ 2 \lambda_{ij,kl}\,  (\varphi_i^\dagger \varphi_j)\,
(\varphi_k^\dagger \nu_l)
+ 2 \lambda_{ij,kl}\,  (\varphi_i^\dagger \varphi_j)\,
(\nu_k^\dagger \varphi_l)
\nonumber\\
&&
+ \lambda_{ij,kl}\, (\varphi_i^\dagger \varphi_j)\,
(\varphi_k^\dagger \varphi_l).
\label{potential_deviated_fields}
\ea
%


Requiring in Eq.~(\ref{potential_deviated_fields})
that the linear terms in $\varphi_i$ vanish,
gives us the stationarity conditions
\be
\left[ \mu_{ij}
+ 2 \lambda_{ij,kl}\, \nu_k^\dagger \nu_l \right]\ \nu_j = 0
\hspace{3cm}(\textrm{for\ } i = 1, \cdots, N).
\label{stationarity_conditions}
\ee
Multiplying by $\nu_i^\dagger$ leads to
\be
\mu_{ij} (\nu_i^\dagger \nu_j) = - 2 \lambda_{ij,kl}\,
(\nu_i^\dagger \nu_j)\, (\nu_k^\dagger \nu_l).
\label{aux_1}
\ee

The value of the potential at a stationary point is found from
Eq.~(\ref{potential_deviated_fields}) by setting all
$\varphi_i=0$.
Using Eq.~(\ref{aux_1}),
this may be written in the following three forms
\ba
V_H^{\rm stationary\ point}
&=&
\mu_{ij}\, (\nu_i^\dagger \nu_j) +
\lambda_{ij,kl}\, (\nu_i^\dagger \nu_j)\, (\nu_k^\dagger \nu_l)
\label{V_SP_mL}
\\
&=&
\frac{1}{2}\, \mu_{ij}\, (\nu_i^\dagger \nu_j)
\label{V_SP_m}
\\
&=&
- \lambda_{ij,kl}\, (\nu_i^\dagger \nu_j)\, (\nu_k^\dagger \nu_l).
\label{V_SP_L}
\ea

\subsection{\label{subsec:vev_structure}A simple basis to study charge breaking in the NHDM}

After spontaneous symmetry breaking the Higgs fields acquire the
vevs
\be
\left( \begin{array}{c} \nu_1^u\\ \nu_1^d \end{array} \right), \ \ \
\left( \begin{array}{c} \nu_2^u\\ \nu_2^d \end{array} \right), \ \ \
\cdots\ \ \
\left( \begin{array}{c} \nu_{N-1}^u\\ \nu_{N-1}^d \end{array} \right), \ \ \
\left( \begin{array}{c} \nu_N^u\\ \nu_N^d \end{array} \right).
\label{vevs-initial}
\ee

An analysis of the potential with such a complicated vev structure
would be too difficult to perform. We will now show how, using the
freedom to choose a basis for the Higgs doublets, one manages to
simplify immensely this study. We start by performing a unitary
transformation on the last two Higgs fields according to
\be
\left(
\begin{array}{c}
\Phi^\prime_{N-1}\\
\Phi^\prime_N
\end{array}
\right)
=
\frac{1}{\sqrt{|\nu_{N-1}^u|^2 + |\nu_N^u|^2 }}
\left(
\begin{array}{cc}
\nu_{N-1}^{u\ \ast} & \nu_N^{u\,\ast}\\
- \nu_N^u & \nu_{N-1}^u
\end{array}
\right)
\left(
\begin{array}{c}
\Phi_{N-1}\\
\Phi_N
\end{array}
\right).
\label{remove-last-up}
\ee
With this transformation, the vevs of the last two fields become
\ba
\langle \Phi^\prime_{N-1} \rangle
&=&
\frac{1}{\sqrt{|\nu_{N-1}^u|^2 + |\nu_N^u|^2 }}
\left(
\begin{array}{c}
|\nu_{N-1}^u|^2 + |\nu_N^u|^2\\
\nu_{N-1}^{u\ \ast} \nu_{N-1}^d + \nu_N^{u\, \ast} \nu_N^d
\end{array}
\right),
\nonumber\\*[2mm]
\langle \Phi^\prime_N \rangle
&=&
\frac{1}{\sqrt{|\nu_{N-1}^u|^2 + |\nu_N^u|^2 }}
\left(
\begin{array}{c}
0 \\
- \nu_N^u \nu_{N-1}^d + \nu_{N-1}^u \nu_N^d
\end{array}
\right),
\ea
respectively. We have thus succeeded in removing the upper component
of the vev of the last Higgs field. Moreover, the upper component of
$\langle \Phi^\prime_{N-1} \rangle$ became real and positive. We can
continue with similar transformations, applied to successive pairs
of Higgs fields, until the corresponding vevs become
\be
\left( \begin{array}{c} \nu_1^u\\ \nu_1^d \end{array} \right), \ \ \
\left( \begin{array}{c} 0 \\ \nu_2^d \end{array} \right), \ \ \
\cdots\ \ \
\left( \begin{array}{c} 0 \\ \nu_{N-1}^d \end{array} \right), \ \ \
\left( \begin{array}{c} 0 \\ \nu_N^d \end{array} \right).
\label{after-up}
\ee
Notice that $\nu_1^u$ and $\nu_i^d$ in
Eq.~(\ref{after-up}) are \textit{not} the same as in
Eq.~(\ref{vevs-initial}),
but rather the values obtained after the successive
transformations of the type shown in Eq.~(\ref{remove-last-up}).
Similarly, we keep the notation for the fields,
although they have been transformed through a series of
basis changes. At the end of the process outlined,
$\nu_1^u$ is real and positive.

We may now repeat the exercise with the lower components. Indeed,
through the transformation
\be
\left(
\begin{array}{c}
\Phi^\prime_{N-1}\\
\Phi^\prime_N
\end{array}
\right)
=
\frac{1}{\sqrt{|\nu_{N-1}^d|^2 + |\nu_N^d|^2 }}
\left(
\begin{array}{cc}
\nu_{N-1}^{d\ \ast} & \nu_N^{d\,\ast}\\
- \nu_N^d & \nu_{N-1}^d
\end{array}
\right)
\left(
\begin{array}{c}
\Phi_{N-1}\\
\Phi_N
\end{array}
\right),
\label{remove-last-down}
\ee
we can change the vevs of the last two Higgs fields in
Eq.~(\ref{after-up}) into
\ba
\langle \Phi^\prime_{N-1} \rangle
&=&
\left(
\begin{array}{c}
0 \\
\sqrt{|\nu_{N-1}^d|^2 + |\nu_N^d|^2 }
\end{array}
\right),
\nonumber\\*[2mm]
\langle \Phi^\prime_N \rangle
&=&
\left(
\begin{array}{c}
0 \\
0
\end{array}
\right),
\label{mag}
\ea
respectively. This eliminates the lower component of $\langle
\Phi^\prime_N \rangle$ and makes the lower component of $\langle
\Phi^\prime_{N-1} \rangle$ real and positive. We may continue with
the other down components, until we reach the following vev
structure
\be
\left( \begin{array}{c} |\nu_1^u|\\ \nu_1^d \end{array} \right), \ \ \
\left( \begin{array}{c} 0 \\ |\nu_2^d| \end{array} \right), \ \ \
\left( \begin{array}{c} 0 \\ 0 \end{array} \right), \ \ \
\cdots\ \ \
\left( \begin{array}{c} 0 \\ 0 \end{array} \right).
\label{vevs-after-basis-transformations}
\ee
Notice that $|\nu_2^d|$,
which is real and positive due to Eq.~(\ref{mag}),
cannot be removed without implying
the appearance of an upper component on the second vev.
We have thus reached a simple but remarkable result.
Indeed, although there are many parameters involved in
the general N-Higgs doublet model,
its vacuum~\cite{clari} structure can be brought into a much simpler form,
through a suitable basis choice.

If, after all these basis transformations, we are left with a vev
structure for which $\nu_1^u\,\neq\,0$, then the vacuum breaks
electric charge. As in the 2HDM, we may now utilize the gauge
freedom in order to bring the vevs into the final form
\cite{change_vevs_phase}
\be \left( \begin{array}{c} \alpha \\ v_{c1}  \end{array} \right), \
\ \ \left( \begin{array}{c} 0 \\ v_{c2} e^{i \delta_c} \end{array}
\right), \ \ \ \left( \begin{array}{c} 0 \\ 0 \end{array} \right), \
\ \ \cdots\ \ \ \left( \begin{array}{c} 0 \\ 0 \end{array} \right),
\label{vevs-final} \ee
where $\delta_c$ is a phase, while $\alpha$, $v_{c1}$, and $v_{c2}$
are positive real numbers. This, then, is the simplest form one can
find for a CB vacuum.

However, we are interested in comparing the value of the potential
at the CB vacuum with its value at the normal vacuum. Therefore, we
must find out what is the form of the most general normal vacuum, in
the basis in which Eq.~(\ref{vevs-final}) is written. Clearly, given
our definition of electric charge, it will have all $\nu_i^u\,=\,0$.
A generic charge-preserving vacuum, then, will have the form
\be
\left( \begin{array}{c}  0 \\ \nu_1^d \end{array} \right), \ \ \
\left( \begin{array}{c} 0 \\ \nu_2^d \end{array} \right), \ \ \
\cdots\ \ \
\left( \begin{array}{c} 0 \\ \nu_{N-1}^d \end{array} \right), \ \ \
\left( \begin{array}{c} 0 \\ \nu_N^d \end{array} \right).
\label{eq:vnorm}
\ee

We emphasise that these $\nu_i^d$ are {\em not} the same as those
appearing in Eq.~(\ref{vevs-initial}). However, we can now apply the
same method we used previously to bring the normal vacuum to a more
manageable form. Through a transformation analogous to that of
Eq.~(\ref{remove-last-down}), we can set the last doublet to zero.
Notice that this basis change does not involve the first two
doublets, so the charge-breaking vacuum structure,
Eq.~(\ref{vevs-final}), remains unaffected. Successive basis
transformations may be applied that do not change Eq.~(\ref{vevs-final})
but set to zero the lower component vevs of Eq.~(\ref{eq:vnorm}), until
one is left with the final normal vacuum structure,
\be \left( \begin{array}{c} 0 \\ \nu_1^d \end{array} \right), \ \ \
\left( \begin{array}{c} 0 \\ \nu_2^d \end{array} \right), \ \ \
\left( \begin{array}{c} 0 \\ |\nu_3^d| \end{array} \right), \ \ \
\left( \begin{array}{c} 0 \\ 0 \end{array} \right) \ \ \ \cdots\ \ \
\left( \begin{array}{c} 0 \\ 0 \end{array} \right).
\label{eq:vevsnorm} \ee
If we try to perform another basis change to set $|\nu_3^d|$ to
zero, we will destroy the simple form for the charge-breaking vevs,
Eq~(\ref{vevs-final}). We denote by the ``B-basis" the basis where
the charge breaking vevs have the simple form of Eq.~(\ref{vevs-final}) and
the normal vacuum vevs, the form given by
Eq.~(\ref{eq:vevsnorm}). The B-basis is appropriate to study
the possibility of charge breaking vacua. This result shows that the study
of charge breaking for an N-doublet potential is reduced to the analysis of the
three-doublet situation.

\subsection{A simple basis to study CP breaking in the NHDM with explicit CP conservation}

We are also interested in the possibility of CP being spontaneously broken
in N-Higgs doublets models. In this case we cannot simply choose the most general
NHDM potential - we must make sure that that potential does not break CP
explicitly. In appendix~A of \cite{GunHab},
Gunion and Haber invoke CPT and $\mathcal{T}^2=1$ (where $\mathcal{T}$ is the
time-reversal operator) to show that:
``The Higgs potential is explicitly CP-conserving if and only if a basis
exists in which all Higgs potential parameters are real''. We therefore consider
one such basis for our NHDM potential: all of its parameters are real and it
explicitly preserves CP.

Given our definition of electric charge, the most general
charge-preserving vacuum (CP violating or not) will be of the same
form as Eq.~(\ref{eq:vnorm}). Our starting point, however, is not
the basis in which we wrote Eq.~(\ref{eq:vnorm}), but rather a generic
basis for which the parameters of the potential are all real. It is
now convenient to ensure that all basis changes do not introduce
complex parameters in the potential. This means that we are
restricted to orthogonal basis transformations. Even with this
restriction we are still able to simplify immensely the
study of the NHDM potential. This is accomplished through two series
of steps:
\begin{enumerate}
%
\item We start with an orthogonal
transformation on the last two Higgs fields according to
\be
\left(
\begin{array}{c}
\Phi^\prime_{N-1}\\
\Phi^\prime_N
\end{array}
\right)
=
\frac{1}{\sqrt{\textrm{Im}^2\left(\nu^d_{N-1}\right) +
\textrm{Im}^2\left(\nu^d_N\right)}}
\left(
\begin{array}{cc}
    - \textrm{Im}\left(\nu^d_{N-1}\right) &
        - \textrm{Im}\left(\nu^d_N\right)\\
    - \textrm{Im}\left(\nu^d_N\right) &
        \textrm{Im}\left(\nu^d_{N-1}\right)
\end{array}
\right)
\left(
\begin{array}{c}
\Phi_{N-1}\\
\Phi_N
\end{array}
\right).
\label{remove-last-down-im}
\ee
This eliminates the imaginary part of the vev of the last Higgs field.
We can continue with similar transformations,
applied to successive pairs of Higgs fields,
until the corresponding vevs reach a structure of the type
\be
\left( \begin{array}{c} 0 \\ \nu^d_1 \end{array} \right), \ \ \
\left( \begin{array}{c} 0 \\
               \textrm{Re}\left(\nu^d_2\right)\end{array} \right), \ \ \
\cdots\ \ \
\left( \begin{array}{c} 0 \\
               \textrm{Re}\left(\nu^d_{N-1}\right) \end{array} \right), \ \ \
\left( \begin{array}{c} 0 \\
               \textrm{Re}\left(\nu^d_N\right) \end{array} \right).
\label{after-up-im}
\ee
Notice that the $\nu_i^d$ in
Eq.~(\ref{after-up-im}) are \textit{not} the initial ones,
but rather the values obtained after the successive
transformations of the type shown in Eq.~(\ref{remove-last-down-im}).
Similarly, we keep the notation for the fields,
although they have been transformed through a series of
basis changes.
After these steps $\textrm{Im}\left(\nu^d_1\right) < 0$.
%
\item We continue with
the orthogonal transformation on the last two Higgs fields
\be
\left(
\begin{array}{c}
\Phi^\prime_{N-1}\\
\Phi^\prime_N
\end{array}
\right)
=
\frac{1}{\sqrt{\textrm{Re}^2\left(\nu^d_{N-1}\right) +
\textrm{Re}^2\left(\nu^d_N\right)}}
\left(
\begin{array}{cc}
    \textrm{Re}\left(\nu^d_{N-1}\right) &
        \textrm{Re}\left(\nu^d_N\right)\\
    - \textrm{Re}\left(\nu^d_N\right) &
        \textrm{Re}\left(\nu^d_{N-1}\right)
\end{array}
\right)
\left(
\begin{array}{c}
\Phi_{N-1}\\
\Phi_N
\end{array}
\right).
\label{remove-last-up-re}
\ee
This eliminates the vev $\langle \Phi^\prime_N \rangle$,
simultaneously making the lower component of
$\langle \Phi^\prime_{N-1} \rangle$ real and positive.
We can continue with similar transformations,
applied to successive pairs of Higgs fields,
until the corresponding vevs reach a structure of the type
\be
\left( \begin{array}{c} 0 \\ \nu_1^d \end{array} \right), \ \ \
\left( \begin{array}{c} 0 \\
            \textrm{Re}\left(\nu^d_2\right)\end{array} \right), \ \ \
\left( \begin{array}{c} 0 \\ 0 \end{array} \right), \ \ \
\cdots\ \ \
\left( \begin{array}{c}  0  \\ 0 \end{array} \right), \ \ \
\left( \begin{array}{c}  0  \\ 0 \end{array} \right),
\label{after-down-re}
\ee
where $\textrm{Re}\left(\nu^d_2\right) \geq 0$.
\end{enumerate}

Hence, when the scalar potential conserves CP, it is
possible to choose a basis in which only the first two doublets have
vevs, while keeping all parameters in the potential real. At this
point, we distinguish two physically distinct scenarios. If
$\textrm{Im}\left(\nu^d_1\right)=0$, then the vacuum is a normal one
and preserves CP; if not, it spontaneously breaks that symmetry - we
call it a CP violating (CPV) vacuum.

Let us then suppose we started with a normal vacuum, and that we
employed the basis transformations described above until the vacuum
structure was reduced to Eq.~(\ref{after-down-re}), with real vevs
$\nu_1^d = v_1$ and $\nu_2^d = v_2$. Because the remaining vevs are
real, we can perform a final basis transformation on the first two
fields
\be
\left(
\begin{array}{c}
\Phi^\prime_1\\
\Phi^\prime_2
\end{array}
\right)
=
\frac{1}{\sqrt{v_1^2 + v_2^2 }}
\left(
\begin{array}{cc}
v_1 & v_2\\
- v_2 & v_1
\end{array}
\right)
\left(
\begin{array}{c}
\Phi_1\\
\Phi_2
\end{array}
\right),
\label{Higgs-basis-N}
\ee
bringing their vevs into the form
\be n_1 = \left( \begin{array}{c} 0 \\ v  \end{array} \right), \ \ \
n_2 = \left( \begin{array}{c} 0 \\ 0  \end{array} \right),
\label{S-basis:N}
\ee
where $v= \sqrt{v_1^2 + v_2^2}$, and all remaining doublets are
zero. This is known as the ``Higgs basis'' for the normal vacuum in
the 2HDM~\cite{BS,BLS}.

However, we are interested in comparing the value of the potential
at the normal vacuum with its value at a CPV vacuum. Therefore, we
must find out what is the form of the most general CPV vacuum, in
the basis in which Eq.~(\ref{S-basis:N}) is written and with the definition
of electric charge we have adopted. It will be of the form of
Eq.~(\ref{eq:vnorm}), with {\em new} vevs $\tilde{\nu}_i^d$,
\be \left( \begin{array}{c}  0 \\ \tilde{\nu}_1^d \end{array}
\right), \ \ \ \left( \begin{array}{c} 0 \\ \tilde{\nu}_2^d
\end{array} \right), \ \ \ \cdots\ \ \ \left( \begin{array}{c} 0 \\
\tilde{\nu}_{N-1}^d
\end{array} \right), \ \ \ \left( \begin{array}{c} 0 \\ \tilde{\nu}_N^d
\end{array} \right).
\label{eq:vtil} \ee
Because the normal vacuum has been reduced to the form of
Eq.~(\ref{S-basis:N}), where only the first doublet is different
from zero, we can apply the steps 1 and 2 detailed above for this
new vacuum and again reduce the CPV vacuum to the form
\be s_1 = \left( \begin{array}{c} 0 \\ z_1
\end{array} \right), \ \ \
s_2 = \left( \begin{array}{c} 0 \\
            z_2 \end{array} \right),
\label{S-basis:CPV}
\ee
with all remaining doublets zero. In this equation, $z_1$ and $z_2$ are
complex numbers.
The sequence of steps that led us from Eq.~(\ref{eq:vtil}) to
Eq.~(\ref{S-basis:CPV}) did {\em not} change the normal vacuum
of Eq.~(\ref{S-basis:N}),
because none of those steps involved the first doublet.
We cannot further remove the imaginary or real parts of $z_2$
because that operation would involve the first doublet and, thus,
take us away from a basis in which the form of the normal
vacuum of Eq.~(\ref{S-basis:N}) remains valid.
The basis for which the normal and the CPV stationary
points have the simple form given by Eqs.~(\ref{S-basis:N})
and (\ref{S-basis:CPV}) will be called the ``S-basis".

The S-basis is very useful because, when using it for the normal
minimum, the Goldstone bosons are isolated as the components of
$\varphi_1$, while the other $\varphi_i$ ($i=2, \cdots, N$) contain
other charged and neutral scalars fields \cite{BS,BLS,LS}. Indeed,
in the S-basis
\ba
\varphi_1 &=&
\left( \begin{array}{c}
            G^+\\
            (H^0 + i G^0)/\sqrt{2}
       \end{array}
\right),
\label{phi1_Higgs_basis}
\\
\varphi_i &=&
\left( \begin{array}{c}
            H^+_i\\
            (R_i + i I_i)/\sqrt{2}
       \end{array}
\label{phiothers_Higgs_basis}
\right),\ \ \ \ \ \ (\textrm{for\ } i=2, \cdots, N),
\ea
where $ G^+ $ and $ G^0 $ are the Goldstone bosons (which, in the
unitary gauge, become the longitudinal components of the $ W^+ $ and
of the $ Z^0 $); $ H^0 $ couples to fermions proportionally to their
masses (in the fermion mass basis); and $H^+_i$, $R_i$, and $I_i$
($i=2, \cdots, N$) are the charged and neutral scalars fields.
Notice that these are not the physical particles; those will be
obtained by diagonalizing the squared mass matrix of the charged
Higgs, and the squared mass matrix of the neutral Higgs (including
$H^0$, $R_i$ and $I_i$). These important properties will become
obvious below.

\subsection{\label{subsec:mass_terms}The mass terms
at the normal vacua}

We now wish to study the quadratic terms in
Eq.~(\ref{potential_deviated_fields}), when the vevs are taken to
coincide with those at a normal vacuum. Since the basis
transformations do not mix the upper and lower components, the
normal vacua have $\nu_i^u=0$ for all $i$, in any basis (as long as
no gauge transformations are made). As a result, the quadratic terms
of Eq.~(\ref{potential_deviated_fields}) evaluated at a normal
stationary point may be written, in any basis, as
\ba
\left(M^2\right)^{\rm N}_{ij} &=&
\left(M^2_{\pm}\right)^{\rm N}_{ij}\,
\varphi_i^{u\, \ast}\,
\varphi_j^u
+
\left(M^2_{R}\right)^{\rm N}_{ij}\,
\textrm{Re}\left(\varphi_i^d\right)\,
\textrm{Re}\left(\varphi_j^d\right)
+
\left(M^2_{I}\right)^{\rm N}_{ij}\,
\textrm{Im}\left(\varphi_i^d\right)\,
\textrm{Im}\left(\varphi_j^d\right)
\nonumber\\
&&
+
\left(M^2_{RI}\right)^{\rm N}_{ij}\,
\textrm{Re}\left(\varphi_i^d\right)\,
\textrm{Im}\left(\varphi_j^d\right)
+
\left(M^2_{IR}\right)^{\rm N}_{ij}\,
\textrm{Im}\left(\varphi_i^d\right)\,
\textrm{Re}\left(\varphi_j^d\right),
\label{mass_terms_N}
\ea
where we identify
\ba
\left(M^2_{\pm}\right)^{\rm N}_{ij} &=&
\mu_{ij} + 2 \lambda_{ij,kl} \nu_k^{d\, \ast}\, \nu_l^d,
\label{M_pm_general}
\\
\left(M^2_{R}\right)^{\rm N}_{ij} &=&
\textrm{Re}\left[
\mu_{ij}
+ 2 \lambda_{ij,kl} \nu_k^{d\, \ast}\, \nu_l^d
+ 2 \lambda_{ik,lj} \nu_k^d\, \nu_l^{d\, \ast}
+ 2 \lambda_{ik,jl} \nu_k^d\, \nu_l^d
\right],
\\
\left(M^2_{I}\right)^{\rm N}_{ij} &=&
\textrm{Re}\left[
\mu_{ij}
+ 2 \lambda_{ij,kl} \nu_k^{d\, \ast}\, \nu_l^d
+ 2 \lambda_{ik,lj} \nu_k^d\, \nu_l^{d\, \ast}
- 2 \lambda_{ik,jl} \nu_k^d\, \nu_l^d
\right],
\label{M_I_general}
\\
\left(M^2_{RI}\right)^{\rm N}_{ij} &=&
- \textrm{Im}\left[
\mu_{ij}
+ 2 \lambda_{ij,kl} \nu_k^{d\, \ast}\, \nu_l^d
+ 2 \lambda_{ik,lj} \nu_k^d\, \nu_l^{d\, \ast}
- 2 \lambda_{ik,jl} \nu_k^d\, \nu_l^d
\right],
\\
\left(M^2_{IR}\right)^{\rm N}_{ij} &=& \textrm{Im}\left[ \mu_{ij} +
2 \lambda_{ij,kl} \nu_k^{d\, \ast}\, \nu_l^d + 2 \lambda_{ik,lj}
\nu_k^d\, \nu_l^{d\, \ast} + 2 \lambda_{ik,jl} \nu_k^d\, \nu_l^d
\right],
\label{eq:Mir}
\ea
and the superscript N indicates that these mass matrices have been
evaluated at the normal vacuum.

Using Eqs.~(\ref{hermiticity_coefficients}), one can show that the
matrix $(M^2_{\pm})^{\rm N}$ is hermitian, while the real matrices
$(M^2_{R})^{\rm N}$ and $(M^2_{I})^{\rm N}$ are symmetric.
The remaining two matrices are real and related by
$\left(M^2_{RI}\right)^{\rm N}_{ji} = \left(M^2_{IR}\right)^{\rm
N}_{ij}$. This implies that the $2N \times 2N$ matrix,
\be
\left(
\begin{array}{cc}
(M^2_{R})^{\rm N} & (M^2_{RI})^{\rm N}
\\
(M^2_{IR})^{\rm N} & (M^2_{I})^{\rm N}
\end{array}
\right),
\label{huge_M}
\ee
is symmetric.
Moreover, the matrix $(M^2_{\pm})^{\rm N}$ behaves like
a second rank tensor under a basis transformation of the Higgs
fields, but the other matrices do not.

As we mentioned, these expressions are valid for normal vacua in any
basis.
For the B-basis,
where only $\{\nu_1^d\,,\,\nu_2^d\,,\,\nu_3^d\}$ are different
from zero,
the indices $\{k\,,\,l\}$ in the
Eqs.~(\ref{M_pm_general})-(\ref{eq:Mir}) run only from 1 to 3.
In the S-basis, where only the first doublet has a non-zero vev, the
mass matrices are simplified considerably. In what follows the vevs,
the parameters $\mu_{ij}$, and $\lambda_{ij,kl}$ are all written in
the S-basis; it is important to understand that changing the basis
would change the vevs, but also the parameters $\mu_{ij}$ and
$\lambda_{ij,kl}$ \cite{BS}. We find,
\ba
\left(M^2_{\pm}\right)^{\rm N}_{ij} &=& \mu_{ij} + 2 v^2
\lambda_{ij,11}, \label{M_pm}
\\
\left(M^2_{R}\right)^{\rm N}_{ij} &=&
\textrm{Re}\left[
\mu_{ij} + 2 v^2
\left( \lambda_{ij,11} + \lambda_{i1,1j} + \lambda_{i1,j1} \right)
\right],
\\
\left(M^2_{I}\right)^{\rm N}_{ij} &=&
\textrm{Re}\left[
\mu_{ij} + 2 v^2
\left( \lambda_{ij,11} + \lambda_{i1,1j} - \lambda_{i1,j1} \right)
\right],
\label{M_I}
\\
\left(M^2_{RI}\right)^{\rm N}_{ij} &=&
- \textrm{Im}\left[
\mu_{ij} + 2 v^2
\left( \lambda_{ij,11} + \lambda_{i1,1j} - \lambda_{i1,j1} \right)
\right],
\\
\left(M^2_{IR}\right)^{\rm N}_{ij} &=&
\textrm{Im}\left[
\mu_{ij} + 2 v^2
\left( \lambda_{ij,11} + \lambda_{i1,1j} + \lambda_{i1,j1} \right)
\right].
\ea
Using the parametrization of the normal stationary point in the
S-basis, shown in Eq.~(\ref{S-basis:N}), on the stationarity
conditions of Eq.~(\ref{stationarity_conditions}), we find
\be
\mu_{i1} + 2 v^2 \lambda_{i1,11} = 0.
\label{SC_N_Sbasis}
\ee
But this coincides with the definition of $(M^2_{\pm})^{\rm
N}_{i1}$, in Eq.~(\ref{M_pm}) and, since this is a hermitian matrix,
we conclude that the first row and the first column of
$(M^2_{\pm})^{\rm N}$ have zero in every entry,
\be
(M^2_{\pm})^{\rm N}_{i1} = 0 = (M^2_{\pm})^{\rm N}_{1i}
\hspace{2cm} (\textrm{for } i=1, \cdots, N).
\label{zeros_of_Mpm}
\ee
This shows that, indeed, $\varphi^u_1$ in the S-basis coincides with
the charged Goldstone boson, in accordance with
Eq.~(\ref{phi1_Higgs_basis}).

Also,
using Eq.~(\ref{M_I}) with $j=1$ and Eq.~(\ref{SC_N_Sbasis}),
\be
\left(M^2_{I}\right)^{\rm N}_{i1} =
\textrm{Re}\left( \mu_{i1} + 2 v^2 \lambda_{i1,11} \right)
+ 2 v^2 \textrm{Re}\left(\lambda_{i1,11} - \lambda_{i1,11} \right)
= 0.
\ee
Since $\left(M^2_{I}\right)^{\rm N}$ is symmetric,
we find
\be
(M^2_I)^{\rm N}_{i1} = 0 = (M^2_I)^{\rm N}_{1i}
\hspace{2cm} (\textrm{for } i=1, \cdots, N).
\label{zeros_of_MI}
\ee
To simplify, let us now consider for a moment the case in which all
$\mu_{ij}$ and all $\lambda_{ij,kl}$ are real. As we explained
above, a CP-conserving NHDM potential falls under this category. In
that case, $(M^2_{RI})^{\rm N}=0=(M^2_{IR})^{\rm N}$, and the matrix
in Eq.~(\ref{huge_M}) becomes block diagonal. In addition,
$\left(M^2_{R}\right)^{\rm N}$ and $\left(M^2_{I}\right)^{\rm N}$
are the squared mass matrices of the scalars and pseudoscalars,
respectively. Thus, Eq.~(\ref{zeros_of_MI}) shows that in
the S-basis $\textrm{Im}(\varphi^d_1)$ coincides with the neutral,
pseudoscalar Goldstone boson, in accordance with
Eq.~(\ref{phi1_Higgs_basis}).

\section{\label{sec:CB}The charge breaking versus the
normal stationary points}

Throughout this section we will work in the B-basis, although it
will be obvious that our final results hold in any basis. We assume
that both the N and CB stationary points exist. We will now compute
the values of the potential at each of those stationary points. To
that effect, we first recall the results of
section~\ref{sec:N-Higgs}, where we showed that, in the B-basis, the
vev structure of both stationary points is given by
\ba
c_1 = \left( \begin{array}{c} \alpha \\ v_{c1}  \end{array}
\right),
\ \ \
c_2 = \left( \begin{array}{c} 0 \\ v_{c2} e^{i
\delta_c} \end{array} \right),
\ \ \
c_3 = \left( \begin{array}{c} 0
\\ 0 \end{array} \right),
&&\ \ \ \ \ \textrm{Charge\ breaking\ vacuum\ (CB)},
\nonumber\\*[3mm]
 n_1 = \left(
\begin{array}{c} 0 \\ \nu_1  \end{array} \right),
\ \ \
n_2 = \left(
\begin{array}{c} 0 \\ \nu_2  \end{array} \right),
\ \ \
n_3 = \left(
\begin{array}{c} 0 \\ v_3  \end{array} \right),
&&\ \ \ \ \  \textrm{Normal\ vacuum\ (N)}, \label{vevsbb} \ea
with all remaining doublets having vevs equal to zero.
The parameter $\delta_c$ is a phase, while $\alpha$, $v_{ci}$ and $v_3$ are real,
positive numbers.
In general, $\nu_1$ and $\nu_2$ are complex.
From Eq.~(\ref{stationarity_conditions}), we obtain the stationarity
conditions for the CB stationary point in the B-basis,
\be \left(\mu_{i1} + 2 \lambda_{i1,kl}\, c_k^\dagger
c_l\right)\,\alpha \, = \,0 \label{eq:al} \ee
\be \left(\mu_{i1} + 2 \lambda_{i1,kl}\, c_k^\dagger
c_l\right)\,v_{c1} \,+\, \left(\mu_{i2} + 2 \lambda_{i2,kl}\,
c_k^\dagger c_l\right)\,v_{c2} e^{i \delta_c}\, = \,0 \;\;\; , \label{eq:vs}
\ee
for $i\,=\,1\,\cdots\,,\,N$ and $k,l \,=\,1\,,\,2$. Since we assume
that the CB stationary point exists, $\alpha\neq 0$ and its
coefficient in Eq.~(\ref{eq:al}) must equal zero. From
Eq.~(\ref{eq:vs}), then, the coefficient of $v_{c2} e^{i \delta_c}$
is also zero. As a result, the stationarity conditions at the CB
stationary point may be written as
\be
\mu_{ij_1} + 2 \lambda_{ij_1,kl}\, c_k^\dagger c_l = 0
\hspace{2cm} (i= 1 \, \cdots, N\, ; \ \ j_1=1,2).
\label{SC_CB_Sbasis_2}
\ee

Let us now contract the indices $\{i\,,\,j_1\}$ with $n_i^\dagger
n_{j_1}$. This gives
\be \left(\mu_{ij_1} + 2 \lambda_{ij_1,kl}\, c_k^\dagger
c_l\right)\,n_i^\dagger n_{j_1}  = 0 \hspace{2cm} (i= 1, 2, 3 \,;\ \
j_1 = 1,2), \ee
and from here it is trivial to obtain
\be \mu_{ij}\,n_i^\dagger n_j + 2 \lambda_{ij,kl}\, n_i^\dagger n_j
c_k^\dagger c_l - \left(\mu_{i3} + 2 \lambda_{i3,kl}\, c_k^\dagger
c_l\right)\,n_i^\dagger n_3   = 0 \hspace{2cm} (i,j = 1 \, \cdots,
3). \label{eq:intcb} \ee
Notice the appearance of the term $\mu_{ij}\,n_i^\dagger n_j$ which,
according to Eq.~(\ref{V_SP_m}), equals twice the value of of the
potential at the normal stationary point, $V_H^N$.

Now, from Eq.~(\ref{M_pm_general}), the mass matrix for the charged
scalars at the N vacuum in the B-basis is given by
\ba \left(M^2_{\pm}\right)^{\rm N}_{kl} &=& \mu_{kl} + 2
\lambda_{kl,ij} n_i^\dagger n_j \;\;\; .
\ea
Contracting the indices $\{k\,,\,l\}$ with $c_k^\dagger c_l$ we
obtain
\begin{align}
\left(M^2_{\pm}\right)^{\rm N}_{kl}\,c_k^\dagger c_l &=
\;\;\;\mu_{kl}\,c_k^\dagger c_l \,+\, 2 \lambda_{kl,ij}\,
n_i^\dagger n_j \, c_k^\dagger c_l
\nonumber \\
 &= \;\;\;2 V_H^{CB} + 2 \lambda_{ij,kl}\,
n_i^\dagger n_j\,  c_k^\dagger c_l \;\;\;,
\label{eq:intn}
\end{align}
where we have used Eq.~(\ref{V_SP_m}) to identify
$\mu_{kl}\,c_k^\dagger c_l$ as twice the value of the potential at
the CB stationary point and the symmetries of the $\lambda$
coefficients from Eq.~(\ref{hermiticity_coefficients}). Comparing
Eqs.~(\ref{eq:intcb}) and (\ref{eq:intn}), we can subtract them to
find
\begin{equation}
V_H^{CB} \,-\, V_H^N \;=\; \frac{1}{2}\left(M^2_{\pm}\right)^{\rm
N}_{ij}\,c_i^\dagger c_j \,-\,\frac{1}{2}\, v_3\,\left(\mu_{i3} + 2
\lambda_{i3,kl}\, c_k^\dagger c_l\right)\,n_i\;\;\;.
\label{eq:vcbvn}
\end{equation}
This is our main result regarding the possibility of charge breaking
in the NHDM. Although obtained in the B-basis, it is very simple to
rewrite Eq.~(\ref{eq:vcbvn}) in a basis invariant form. At this point it
is important to recall the results obtained in reference~\cite{BFS}
for CB in the case of the 2HDM. In the notation of this paper, the
conclusions therein reached are written as
\begin{equation}
V_H^{CB} \,-\, V_H^N \;=\; \frac{1}{2}\left(M^2_{\pm}\right)^{\rm
N}_{ij}\,c_i^\dagger c_j \;\;\; .
\end{equation}
When the normal stationary point is a minimum, the matrix
$\left(M^2_{\pm}\right)^{\rm N}$ has, besides the Goldstone bosons,
only positive eigenvalues, and it is very easy to prove~\cite{BFS}
that one obtains $V_H^{CB} \,-\, V_H^N \,>\,0$. Hence, if a
normal minimum exists, the CB stationary point is {\em always} above
it. No possibility of tunneling from the normal minimum to a deeper
CB stationary point exists in the 2HDM.

The similarity with the NHDM case is clear, but the difference of the potential
depths now contains an extra term, proportional
to $v_3$. Let us consider that the normal vacuum in the NHDM is
indeed a minimum. Then, as before, the term
$\left(M^2_{\pm}\right)^{\rm N}_{ij}\,c_i^\dagger c_j$ is strictly
positive \cite{demo}.
However, there is no {\em a priori} reason for the second term in the
right-hand side of Eq.~(\ref{eq:vcbvn}) to be positive. And in fact,
depending on the values of the parameters $\mu$ and $\lambda$, it
may well be negative, so much so that it overwhelms the positive
contributions from $\left(M^2_{\pm}\right)^{\rm N}_{ij}\,c_i^\dagger
c_j$.

As an example of this possibility, we undertook a study of CB in the
3HDM for generic values of the parameters of the potential. For simplicity we
considered the 3HDM potential without explicit CP violation. Our
conclusions are as follows:
\begin{enumerate}
\item As in the case of the 2HDM, it is certainly possible to find
combinations of $\{\mu\,,\,\lambda\}$ for which there are normal
minima with a CB stationary point located {\em above} them. 
\item However, unlike the 2HDM situation, we have found combinations
of $\{\mu\,,\,\lambda\}$ for which both the normal and Charge
Breaking stationary points are minima, {\em but} verify
$V_H^{CB}\,<\,V_H^N$.
\end{enumerate}
In Appendix~\ref{ap:3hdm} we give a set of numerical values of
$\{\mu\,,\,\lambda\}$ corresponding to this situation. In fact we obtain, from such
parameter values,
\begin{equation}
V_H^{CB}\,=\,-2.6678 \times 10^9\;\; \mbox{GeV}^4
\,<\,V_H^N\,=\,-2.2792 \times 10^9 \;\; \mbox{GeV}^4 \;\;\; .
\end{equation}
A numerical minimization of the potential found
no value below $V_H^{CB}$.

To ensure that both CB and N are minima, we calculated the scalar
squared mass matrices at both stationary points. Other than the
expected zero eigenvalues (3 for the N minimum, 4 for the CB one)
all the others are positive.

In conclusion, the study of charge breaking vacua in the NHDM
reduces itself to the study of a 3HDM potential. And, for this one -
and unlike the 2HDM case - there is the possibility of CB minima
which are {\em deeper} than a normal minimum.

\section{\label{sec:CPV}The CPV versus the normal stationary points}

Let us consider a Higgs potential with explicit CP conservation and
no CB stationary points. As we showed in section~\ref{sec:N-Higgs},
it is possible, through a series of orthogonal transformations that
preserve $\mu$ and $\lambda$ as real (though changing its values),
to reach what we called the S-basis, where a normal (CB and CP
preserving) and CPV (CB conserving, CP violating) stationary points
have vevs given by
\ba s_1 = \left( \begin{array}{c} 0 \\ z_1  \end{array} \right), \ \
\ s_2 = \left( \begin{array}{c} 0 \\ z_2
\end{array} \right), \nonumber
&&\ \ \ \ \ \ \ \ \ \ \textrm{CP-violating\ vacuum\ (CPV)}, \\*[3mm]
n_1 = \left(
\begin{array}{c} 0 \\ v  \end{array} \right),
\ \ \ n_2 = \left(\begin{array}{c} 0 \\ 0  \end{array} \right), && \
\ \ \ \ \ \ \ \ \ \textrm{Normal\ vacuum\ (N)} \ea
where $v$ is real and at least one of $\{z_1\,,\,z_2\}$ is complex.
Unlike the CB case, we are able to reduce the
study of CP violation in the NHDM to the analysis of only two
doublets. Throughout this section we will work in the S-basis.
We now assume that both the N and CPV stationary points exist.
Eq.~(\ref{SC_N_Sbasis}) shows the stationarity conditions of
Eq.~(\ref{stationarity_conditions}) applied to the normal stationary
point and written in the S-basis. Similarly, using the
parametrization of the CPV vevs in the S-basis, shown in
Eq.~(\ref{S-basis:CPV}), on the stationarity conditions of
Eq.~(\ref{stationarity_conditions}), we find
\be
\left(
\mu_{i1} + 2 \lambda_{i1,kl}\, s_k^\dagger s_l
\right)\,
z_1
+
\left(
\mu_{i2} + 2 \lambda_{i2,kl}\, s_k^\dagger s_l
\right)\,
z_2
=
0.
\label{SC_CPV_Sbasis_1}
\ee
Specifying for $i=1$ and rearranging the terms,
we obtain
\be
z_1\, \mu_{11} + z_2\, \mu_{12}
=
- 2 \lambda_{11,kl}\, s_k^\dagger s_l\ z_1
- 2 \lambda_{12,kl}\, s_k^\dagger s_l\ z_2,
\label{SC_CPV_Sbasis_2}
\ee
This can be viewed as a system of one complex (two real)
equations in the two real unknowns $\mu_{11}$, and $\mu_{12}$.
The solutions are easily obtained.
One finds that
\ba
- \frac{1}{2} \mu_{11}
&=&
\lambda_{11,11} |z_1|^2
+
\lambda_{11,21} \left( z_1^\ast z_2+ z_2^\ast z_1 \right)
+
\left(
\lambda_{11,22} - \lambda_{12,12} + \lambda_{12,21}
\right)
|z_2|^2
\nonumber\\
&=&
\lambda_{11,11} |z_1|^2
+
\lambda_{12,11} z_1^\ast z_2
+
\lambda_{21,11} z_2^\ast z_1
+
\left(
\lambda_{22,11} + \lambda_{21,12} - \lambda_{21,21}
\right)
|z_2|^2
\nonumber\\
&=&
\left( \lambda_{ij,11} + \lambda_{i1,1j} - \lambda_{i1,j1} \right)\,
s_i^\dagger s_j.
\ea
In addition,
the stationarity condition at the normal minimum,
Eq.~(\ref{SC_N_Sbasis}),
yields
\ba
- \frac{1}{2} \mu_{11}
&=& v^2 \lambda_{11,11} = \lambda_{ij,11} n^\dagger_i n_j
\nonumber\\
&=&
\left( \lambda_{ij,11} + \lambda_{i1,1j} - \lambda_{i1,j1} \right)\,
n_i^\dagger n_j.
\ea
We conclude that
\be
\left( \lambda_{ij,11} + \lambda_{i1,1j} - \lambda_{i1,j1} \right)\,
s_i^\dagger s_j
=
\left( \lambda_{ij,11} + \lambda_{i1,1j} - \lambda_{i1,j1} \right)\,
n_i^\dagger n_j.
\label{aux_crucial_CPV_N}
\ee

We are now ready to calculate the difference between the value of
the scalar potential at the CPV stationary point and the value of
the scalar potential at the N stationary point. We start from the
definition of the pseudoscalar mass matrix $(M^2_I)^{\rm N}$ in
Eq.~(\ref{M_I}) and multiply it, respectively, by $s_i^\dagger s_j$
and $n_i^\dagger n_j$, to find
\ba
\frac{1}{2}
\left(M^2_I\right)^{\rm N}_{ij} s_i^\dagger s_j &=&
\frac{1}{2} \mu_{ij}\, s_i^\dagger s_j
+
v^2\,
\left( \lambda_{ij,11} + \lambda_{i1,1j} - \lambda_{i1,j1} \right)\,
s_i^\dagger s_j,
\nonumber\\
\frac{1}{2}
\left(M^2_I\right)^{\rm N}_{ij} n_i^\dagger n_j &=&
\frac{1}{2} \mu_{ij}\,  n_i^\dagger n_j
+
v^2\,
\left( \lambda_{ij,11} + \lambda_{i1,1j} - \lambda_{i1,j1} \right)\,
n_i^\dagger n_j.
\label{aux_CPV_for_diff}
\ea
Subtracting both lines we find
\be
V_H^{\rm CPV} - V_H^{\rm N} = \frac{1}{2}
\left(M^2_I\right)^{\rm N}_{ij} s_i^\dagger s_j.
\label{V_CPV--V_N}
\ee
In obtaining this result we have used Eq.~(\ref{aux_crucial_CPV_N}),
and we noticed that, according to Eq.~(\ref{V_SP_m}),
\ba
V_H^{\rm CPV} &=&
\frac{1}{2} \mu_{ij}\,  s_i^\dagger s_j,
\nonumber\\
V_H^{\rm N} &=&
\frac{1}{2} \mu_{ij}\,  n_i^\dagger n_j.
\ea
Furthermore, we used the fact that Eqs.~(\ref{S-basis:N}) and
(\ref{zeros_of_MI}) imply that
\be
\left(M^2_I\right)^{\rm N}_{ij} n_i^\dagger n_j = 0.
\ee
Eq.~(\ref{V_CPV--V_N}) is the generalization of the results
obtained in ref.~\cite{BFS} for CP violation in the 2HDM.

It can be shown, using the general definition of
$\left(M^2_I\right)^{\rm N}$ in Eq.~(\ref{M_I_general}), that
Eq.~(\ref{V_CPV--V_N}) is invariant under orthogonal basis
transformations, so that in Eq.~(\ref{V_CPV--V_N}) we can actually
consider the indices $\{i\,,\,j\}$ going from 1 to N. For
simplicity, we evaluate it in the S-basis. As before, when N is
a minimum, we will have
\be
V_H^{\rm CPV} - V_H^{\rm N} = \frac{1}{2}
\left(M^2_I \right)^{\rm N}_{22} |z_2|^2
> 0.
\label{V_CPV--V_N_S-basis}
\ee
The result is strictly positive because the only zero eigenvalue of
the matrix $\left(M^2_I\right)^{\rm N}$ is in the first line/row;
the remaining sub-matrix is definite positive. This implies that all
of the elements of its diagonal - such as $\left(M^2_I \right)^{\rm
N}_{22}$ - are positive. This is another advantage of utilizing the
Higgs basis.

Eq.~(\ref{V_CPV--V_N_S-basis}) generalizes the results obtained in
Ref.~\cite{BFS} for the particular case of $N=2$: whenever a normal
minimum exists it is certainly deeper than any CPV stationary point.

Notice that one can obtain a CPV stationary point which is
deeper than a normal stationary point N.
This occurs for parameters such that
$\left(M^2_I\right)^{\rm N}_{22} < 0$.
However, in that case N is {\textit not} a minimum
(although it is a stationary point).

\section{\label{sec:conclusions}Conclusions}

We have studied the vacuum structure of the most general
N-Higgs-doublet model.
We have shown that, in order to compare the depth of the potential
at a normal minimum with its depth at a CB stationary point,
a basis may be chosen such that the vacuum structure mimics that
of the 3HDM.
Similarly,
in order to compare the depth of the potential
at a normal minimum with its depth at a CPV stationary point,
a basis may be chosen such that the vacuum structure mimics that
of the 2HDM.

This great simplification allowed us to generalize the results of
\cite{BFS}, showing that, whenever a normal minimum exists, it is
certainly deeper than any CPV stationary point. However, we found
one remarkable difference regarding CB: whereas in the 2HDM it is
impossible to find CB minima below normal ones, that does not happen
for the NHDM, with $N\geq 3$. This raises the possibility of finding
charge breaking bounds~\cite{frere} for these potentials, which
might improve their predictive power. Notice, however, that if the
parameters of the potential are such that at the N minimum (in the
B-basis) one has $v_3\,=\,0$, one recovers the 2HDM result for the
NHDM potential: if such a normal minimum exists, it is certainly
deeper than the CB one. This can be used as a sufficient condition
to prevent CB from occurring in the NHDM.

It is interesting to note that Eq.~(\ref{eq:vcbvn}) shows that the
difference between the value of the potential at the CB stationary
point and the value of the potential at the normal stationary point
is related to the charged Higgs squared mass matrix. That relation
is perfect for the 2HDM, but ``spoiled" by the $v_3$ terms in
Eq.~(\ref{eq:vcbvn}) for the NHDM. Similarly, when the potential
conserves CP, Eq.~(\ref{V_CPV--V_N}) shows that the difference
between the value of the potential at the CPV stationary point and
the value of the potential at the normal stationary point is
related to the pseudoscalar squared mass matrix. Thus, the depth of
a potential at a stationary point that breaks a given symmetry,
relative to the normal minimum, is related to the squared mass
matrix of the scalar particles directly linked with that symmetry.

\begin{acknowledgments}
This work was supported by the Portuguese \textit{Funda\c{c}\~{a}o para
a Ci\^{e}ncia e a Tecnologia} (FCT) under the contracts
PDCT/FP/FNU/50155/2003,
POCI/FIS/59741/2004,
and CFTP-Plurianual (777).
In addition, P.\ M.\ F.\ is supported
by FCT under contract SFRH/BPD/5575/2001,
and J.\ P.\ S.\ is supported in part by project POCTI/37449/FNU/2001,
approved by the Portuguese FCT and POCTI,
and co-funded by FEDER.
\end{acknowledgments}

\appendix
\section{\label{app:CB}Charge breaking - the kinetic terms}

In this appendix we present a basis and gauge independent definition of the
charge breaking vacua. This could be done by looking at the
Goldstone bosons in the scalar mass matrix, but it is easier to look
at the mass of the photon instead. The kinetic terms for the scalar
fields are
\be
\left|
\left(
i \partial_\mu - \frac{g}{2} \tau_a {W}^a_\mu -
\frac{g^\prime}{2} B_\mu
\right) \Phi_i
\right|^2,
\ee
where $W_\mu^a$ ($a=1,2,3$) and $B_\mu$ are the SU(2)$_L$ and
U(1)$_Y$ gauge bosons, respectively, and  $\tau_a$ ($a=1,2,3$) are
the Pauli matrices. After spontaneous symmetry breaking we obtain
mass terms for the gauge fields given by
\be
\frac{1}{4}
\left|
\left[
\begin{array}{cc}
g^\prime B_\mu + g W^3_\mu  & \sqrt{2} g W^p_\mu\\
\sqrt{2} g W^m_\mu & g^\prime B_\mu - g W^3_\mu
\end{array}
\right]
\left(
\begin{array}{c}
\nu_i^u\\
\nu_i^d
\end{array}
\right)
\right|^2,
\ee
where
\ba
W_\mu^p &=&
\frac{W_\mu^1 - i W_\mu^2}{\sqrt{2}},
\nonumber\\
W_\mu^m &=&
\frac{W_\mu^1 + i W_\mu^2}{\sqrt{2}}.
\ea
After some reorganization, the result is proportional to
\be
\left(
\begin{array}{cccc}
W^m_\mu, & W^p_ \mu, & W^3_ \mu, & \frac{g^\prime}{g} B_\mu
\end{array}
\right)\
\mathcal{M}^{\rm GB}_i\
\left(
\begin{array}{c}
W^{p\, \mu}\\
W^{m\, \mu}\\
W^{3\, \mu}\\
\frac{g^\prime}{g} B^\mu
\end{array}
\right),
\ee
where
\be
\mathcal{M}^{\rm GB}_i =
\left[
\begin{array}{cccc}
|\nu_i^u|^2 + |\nu_i^d|^2 & 0 & 0 & \sqrt{2} \nu_i^{d\, \ast} \nu_i^u\\
0 & |\nu_i^u|^2 + |\nu_i^d|^2 & 0 & \sqrt{2} \nu_i^{u\, \ast} \nu_i^d\\
0 & 0 & |\nu_i^u|^2 + |\nu_i^d|^2 & |\nu_i^u|^2 - |\nu_i^d|^2\\
\sqrt{2} \nu_i^{u\, \ast} \nu_i^d & \sqrt{2} \nu_i^{d\, \ast} \nu_i^u &
         |\nu_i^u|^2 - |\nu_i^d|^2 & |\nu_i^u|^2 + |\nu_i^d|^2
\end{array}
\right],
\ee
for each doublet $\Phi_i$ ($i=1, \cdots N$). Summing over all
doublets and introducing the complex vectors
\ba
z_u = \left\{ \nu_1^u, \nu_2^u, \cdots \nu_N^u \right\},
\nonumber\\
z_d = \left\{ \nu_1^d, \nu_2^d, \cdots \nu_N^d \right\},
\ea
the mass matrix may be written as
\be
\mathcal{M}^{\rm GB}
=
\sum_{i=1}^N \mathcal{M}^{\rm GB}_i
=
\left[
\begin{array}{cccc}
|z_u|^2 + |z_d|^2 & 0 & 0 & \sqrt{2} z_d \textbf{.} z_u\\
0 & |z_u|^2 + |z_d|^2 & 0 & \sqrt{2} z_u \textbf{.} z_d\\
0 & 0 & |z_u|^2 + |z_d|^2  & |z_u|^2 - |z_d|^2 \\
\sqrt{2} z_u \textbf{.} z_d & \sqrt{2} z_d \textbf{.} z_u&
         |z_u|^2 - |z_d|^2 & |z_u|^2 + |z_d|^2
\end{array}
\right],
\ee
with the notation
\be
(z_u \textbf{.} z_d)^\ast =
z_d \textbf{.} z_u \equiv \sum_{i=1}^N \left(\nu_i^d\right)^\ast \nu_i^u.
\label{define-dot}
\ee
The determinant becomes
\be \textrm{det}\ \mathcal{M}^{\rm GB} = 4 \left(|z_u|^2 + |z_d|^2
\right)^2 \left[ |z_u|^2  |z_d|^2 - |z_d \textbf{.} z_u|^2 \right].
\label{align} \ee

This allows us to define the charge-breaking (CB) vacuum in a
completely basis and gauge independent fashion. Indeed, in order for the
vacuum to conserve charge, \textit{i.e.}, to conserve U(1)$_{\rm
elmg}$, we need to have a massless photon. But that implies that the
determinant in Eq.~(\ref{align}) must vanish. Any combination of
vevs $\{\nu_i^u\,,\,\nu_i^d\}$ for which this does not occur is
therefore a CB stationary point.
As is easily seen from Eq.~(\ref{align}),
we cannot get a CB vacuum with only one Higgs doublet,
a well known result.

\section{\label{ap:3hdm} A three Higgs doublet potential without explicit CP violation}

In this appendix we give a specific example of a 3HDM potential
without explicit CP breaking for which one finds a CB minimum deeper
than a normal one. Since no explicit CP breaking occurs, we work in
a basis where all $\{\mu\,,\,\lambda\}$ are real. The values of the
parameters are given below.

\begin{table}[h]
\caption{Values of the $\mu$ parameters (GeV$^2$).}
\begin{ruledtabular}
\begin{tabular}{cccccc}
$\mu_{11}$ & $\mu_{12}$ & $\mu_{13}$ & $\mu_{22}$ & $\mu_{23}$ &
$\mu_{33}$
\\
\hline
$-7.0655 \times 10^4$ &
$1.6359 \times 10^4$ &
$-2.0184
\times 10^4$ &
$-1.2587 \times 10^4$ &
$-1.7382 \times 10^4$ &
$ 5.0687 \times 10^4$ \\
\end{tabular}
\end{ruledtabular}
\end{table}

\begin{table}[h]
\caption{Values of the $\lambda$ parameters.}
\begin{ruledtabular}
\begin{tabular}{ccccccccc}
$\lambda_{11,11}$ & $\lambda_{11,12}$ & $\lambda_{11,13}$ &
$\lambda_{11,22}$ & $\lambda_{11,23}$ & $\lambda_{11,33}$ &
$\lambda_{12,12}$ & $\lambda_{12,13}$ & $\lambda_{12,21}$ \\
\hline 0.6385 & -0.4227 & -0.0347 & 0.2500 & 0.3128 & 0.8696 &
-0.0987 & 0.2285 & 0.3917 \\
 & & & & & & & & \\
$\lambda_{12,22}$ & $\lambda_{12,23}$ & $\lambda_{12,31}$ &
$\lambda_{12,32}$ & $\lambda_{12,33}$ & $\lambda_{13,13}$ &
$\lambda_{13,22}$ & $\lambda_{13,23}$ & $\lambda_{13,31}$ \\ \hline
-0.3132 & 0.1735 & 0.1780 & -0.0190 & 0.4370 & 0.1852 & -0.1830 &
0.3268 & 0.1620 \\
 & & & & & & & & \\
$\lambda_{13,32}$ & $\lambda_{13,33}$ & $\lambda_{22,22}$ &
$\lambda_{22,23}$ & $\lambda_{22,33}$ & $\lambda_{23,23}$ &
$\lambda_{23,32}$ & $\lambda_{23,33}$ & $\lambda_{33,33}$ \\
\hline -0.1566 & -0.2230 & 0.2373 & -0.2803 & -0.1203 & 0.0536 &
0.4147 & 0.1545 & 0.5368
\\
\end{tabular}
\end{ruledtabular}
\end{table}
All remaining parameters are obtained from these using the
symmetries of $\{\mu\,,\,\lambda\}$ expressed in
Eq.~(\ref{hermiticity_coefficients}). As mentioned this set of
parameters gives us a normal minimum and a CB one. The values of the
vevs obtained are given below (we considered a CB minimum with the
phase $\delta_c$ equal to zero).
\begin{table}[h]
\caption{Values of the vevs for the normal and CB minima (GeV).}
\begin{ruledtabular}
\begin{tabular}{cccccc}
$v_1$ & $v_2$ & $v_3$ & $v_{c1}$ & $v_{c2}$ & $\alpha$ \\ \hline
225.2135 & -41.9564 & 89.6355 &
 325.5199 & 343.9166 & 17.9887 \\
\end{tabular}
\end{ruledtabular}
\end{table}
With these vevs one finds the values of the potential quoted in the
main text. Using the methods developed in the first paper
of~\cite{BFS} it is a simple matter to write down the squared mass
matrices for the scalar fields. They are found to to have, other
than the expected Goldstone bosons, only positive eigenvalues for
both the CB and N stationary points, thus proving that both are
minima.


\begin{thebibliography}{99}
\bibitem{clari}
By ``vacuum" we mean any vev configuration at a stationary point,
not necessarily at a minimum. In fact many of these ``vacua" may well
end up being revealed as saddle points or maxima, as occurs in the
2HDM~\cite{BFS}.
\bibitem{change_vevs_phase}
Since in Eq.~(\ref{vevs-after-basis-transformations})
$\nu_1^u$ and $\nu_2^d$ are real,
we may change the notation into
$\nu_1^u = \alpha$, $\nu_1^d = v_1 e^{- i \delta}$,
and $\nu_2^d = v_2$.
We may then apply an SU(2) gauge transformation combined
with a (hypercharge) U(1) transformation:
%
\be
e^{- i \delta \tau_3 /2} e^{i \delta \mathbf{1} /2}
%
%
=
\left(
\begin{array}{cc}
  1 & 0\\
  0 & e^{i \delta}
\end{array}
\right),
\ee
%
where $\mathbf{1}$ is the $2 \times 2$ unit matrix.
This transforms Eq.~(\ref{vevs-after-basis-transformations})
into Eq.~(\ref{vevs-final}).
\bibitem{demo}
The demonstration is trivial, using the
stationarity conditions for the normal
minimum in Eq.~(\ref{stationarity_conditions}), the form of the CB vevs
given in Eq.~(\ref{vevsbb}) and the definition of
$\left(M^2_{\pm}\right)^{\rm N}$ in Eq.~(\ref{M_pm_general}).
\end{thebibliography}
\end{document}